\documentclass[english]{article}
\usepackage[T1]{fontenc}
\usepackage[latin9]{inputenc}
\usepackage{amsbsy}
\usepackage{amssymb}
\usepackage{esint}

\newcommand{\lyxaddress}[1]{
\par {\raggedright #1
\vspace{1.4em}
\noindent\par}
}

\usepackage{babel}

\begin{document}

\title{Particle Physics challenges to the Bohm Picture of Relativistic Quantum
Field Theory}

\author{A. Miranda}

\maketitle

\lyxaddress{Department of Physics and Astronomy, Aarhus University, DK-8000 Aarhus
C, Denmark}
\begin{abstract}
I discuss selected topics in contemporary Particle Physics from the
point of view of the original ontological formulation of Relativistic
Quantum Field Theory (RQFT), primarily due to D.Bohm and B.Hiley.

The basic platform for doing specific calculations in this paper begins
systematically with modern textbook accounts and techniques. The Bohm-inspired
causal RQFT (referred to as BP) is considered to be, even at this
early stage of its development, a most useful and illuminating $supplement$
to this standard RQFT. 

Unfortunately (in the opinion of this author) it cannot as yet, and
need not, replace the contemporary textbook RQFT. It has - for any
foreseeable future - to work together with it, and thus prepare the
ground for its own future independent developments. Its main role
at this stage is to inspire, to give intuitive, imaginative and creative
$explanations$ of the facts exposed by experimental reserch that
is so excellently $described$ by standard papers and textbooks. It
goes without saying that such fundamental changes in insight could
inspire both theoretical and experimental research to try new paths
that would be otherwise unthinkable. 

I shall try to illustrate these claims by randomly picking up 3 good
examples from contemporary research in Particle Physics, all having
to do with the same basic theme, viz. time development of quantum
systems and particularly, quantum transitions. I work entirely from
the point of view of the orthodox Bohm-Hiley metatheory, but always
trying to make the connection to the present day standard research
that ultimately may qualify to be buried in textbooks.
\end{abstract}

\section{Introduction}

I have recently advertised {[}1{]} for practicing nuclear and particle
theoreticians a novel ontological formulation of non-relativistic
Quantum Mechanics (BP), originally due to de Broglie, but independently
discovered and developed by D. Bohm and his co-workers {[}2{]},{[}3{]},{[}4{]}.
The discussion was illustrated with additional examples taken from
contemporary Nuclear and Particle Physics.

M. R. Brown and B. Hiley discussed not so long ago the interesting
possibility of considering the Heisenberg Picture (HP) as an alternative
starting point {[}5{]} for developing the theory, and how in that
case the original BP could be affected. Further work along these lines
can be found in references {[}6,7{]}. The present paper sticks to
the original Bohm-Hiley version of RQFT {[}2{]} and further developed
by some members of their group,e.g. {[}8{]}, which I shall refer to
as the {}``Bohm school''. A large number of alternative ontological
proposals exist to-day (some representative examples can be found
in references {[}9{]},{[}10{]},{[}11{]},{[}12{]},{[}13{]}).

However, the present paper is not a review article. I feel that these
alternatives have rather different fundamental assumptions, and pursue
different lines of enquiry. They are therefore not considered here. 

A central feature of the Bohm school is described by Bohm himself
in an interview given a long time ago {[}14{]}. It refers to the meaning
he assigns to the word {}``explanation''. This I have systematically
adopted in the following sections. The discussion deals at a certain
stage with the famous {}``gedanken'' two-slit experiments, with
which most of good undergraduate textbooks on Quantum Theory begin.
I quote (my italics):

BOHM: (...) In this way you can $explain$, say, the two-slit experiment.

DAVIES: This is normaly $explained$, of course, by proposing the
interference between waves passing through the two slits.

BOHM: It's not explained, it is merely described. If you said it was
a wave, that would be an explanation. But since the electrons arrive
as particles, it is no explanation. It is a sort of a metaphorical
way of talking. Right? There is no explantion. We should say that
quantum mechanics does not explain anything. It merely gives a formula
for certain results. And I'm trying to give an explanation''.

Be it as it may, the final verdict for that we shall refer to as the
Bohm Picture (BP) (to be added to the standard Schroedinger Picture
(SP), the Heisenberg Picture (HP) and the Dirac Picture (DP)) is not
just around the corner. The main issue for now is rather to patiently
carry on with the nurturing and the step by step testing this kind
of novel and promising attempts, initiated long ago by Bohm's deep
insights into the physical world.

The present particular modest contribution discusses 3 study cases
of interest to the community of practicing particle physicists.

\section{A short review of the Bohm Picture (BP) of Relativistic Quantum Field
Theory (RQFT)}

Let us shortly review some relevant and proeminent features of the
BP, referring the reader to the original literature for full details,
e.g. {[}2{]},{[}3{]},{[}8{]}.

\subsection{Bohm's field equations for Bose-Einstein fields (integer spin fields)}

We begin with Bose-Einstein fields (in modern parlance: integer spin
fields) in the Schroedinger representation:

\begin{equation}
\{\varphi(\vec{x})\}\equiv\varphi_{1}(\vec{x}),\varphi_{2}(\vec{x}),\varphi_{3}(\vec{x}),...\label{eq:1}\end{equation}

They have well-defined classical analogs. A complex field will be
considered as an ordered pair of real fields. The time-dependent Schroedinger
equation is (in units that are used here we assume $\hbar=1=c$, unless
otherwise stated): 

\begin{equation}
H\Psi(\{\varphi(\vec{x})\},t)=i\frac{\partial}{\partial t}\Psi(\{\varphi(\vec{x})\},t)\label{eq:(2)}\end{equation}

where $H$ is the Hamitonian in the Schroedinger representation.

The solutions are defined if some initial condition for the time-dependent
wave-functional is given:

\begin{equation}
\Psi(\{\varphi(\vec{x})\},0)=F(\vec{x},...)\label{eq:(3)}\end{equation}

The Schroedinger representation for field operators is

\begin{equation}
\hat{\varphi}_{k}(\vec{x})|\{\varphi(\vec{x})\}>=\varphi_{k}(\vec{x})|\{\varphi(\vec{x})\}>\label{eq:(4)}\end{equation}

In this representation the canonically conjugate momentum to the field
$\varphi_{k}(\vec{x})$ becomes

\begin{equation}
\hat{\pi}_{k}(\vec{x})\equiv-i\frac{\delta}{\delta\varphi_{k}(\vec{x})}\label{eq:(5)}\end{equation}

We shall consider quantum Hamiltonians of the general form (in Schroedinger
representation)

\begin{equation}
H=\sum_{k}\int d^{3}\vec{x}[-\frac{1}{2}\frac{\delta^{2}}{\delta\varphi_{k}^{2}(\vec{x})}+\frac{1}{2}|\vec{\nabla}\varphi_{k}(\vec{x})|^{2}]+V(\{\varphi(\vec{x})\})\label{eq:(6)}\end{equation}

Let us now begin by briefly reviewing how the BP works in this case.
The full Hamiltonian minus the quadratic part, i.e. the entire non-quadratic
part, will in general be referred to as {}``the interaction Hamiltonian''.
We start as usual {[}1{]},{[}2{]}, by going over to the polar representation
of the wavefunctional:

\begin{equation}
\Psi(\{\varphi(\vec{x})\},t)\equiv R(\{\varphi(\vec{x}),t\},t)e^{iS(\{\varphi(\vec{x})\},t)}\label{eq:(7)}\end{equation}

where R and S are two real wavefunctionals. One inserts this into
the Schroedinger equation (2) and one obtains two coupled partial
differential functional equations:

\begin{equation}
\frac{\partial S}{\partial t}+\frac{1}{2}\sum_{k}\int d^{3}\vec{x}[(\frac{\delta S}{\delta\varphi_{k}(\vec{x})})^{2}+|\vec{\nabla}\varphi_{k}(\vec{x})|^{2}]+V+Q=0\label{eq:(8)}\end{equation}

\begin{equation}
\frac{\partial}{\partial t}P(\{\varphi(\vec{x}),t\})+\sum_{k}\int d^{3}\vec{x}\frac{\delta}{\delta\varphi_{k}(\vec{x})}J_{k}(\{\varphi(\vec{x}),t\})=0\label{eq:(9)}\end{equation}

These two functional equations are part of the basic quantum field
dynamics in BP. However , they are not the complete story (subsection
2.3).

($\mathbf{i}$) The symbol $Q$ stands for the Super-Quantum-Information-Potential
(SQIP) and is the natural field theoretic generalisation of the Bohm
many-body Quantum-Information-Potential (QIP) discussed in {[}1{]},{[}2{]}:

\begin{equation}
Q(\{\varphi(\vec{x})\},t)\equiv-\frac{1}{2}\sum_{k}\int d^{3}\vec{x}\frac{1}{R}\frac{\delta^{2}R(\{\varphi(\vec{x})\},t)}{\delta\varphi_{k}^{2}(\vec{x})}\label{eq:(10)}\end{equation}

We have defined

\begin{equation}
P\equiv P(\{\varphi(\vec{x})\},t)\equiv R^{2}(\{\varphi(\vec{x})\},t)\equiv|\Psi(\{\varphi(\vec{x})\},t)|^{2}\label{eq:(11)}\end{equation}

The functional P has a double role assigned to it, just as it is the
case with the non-relativistic formulation {[}1{]},{[}2{]}. First,
according to equation (9), P is the locally conserved $probability$
that the configuration of the fields $at\: time\: t\; is$ $\{\varphi(\vec{x})\}$
{[}2{]}; this is the role \#1 that we shall assign to $P$ in the
present paper. Next, but not least, role \#2 is that P is of paramount
importance in determining $the$ central quantity in Bohm's ontological
formulation of Quantum Theory, i.e the SQIP itself, definition (10){[}2{]}.
We shall nevertheless not insist on this most important feature of
P that is a direct consequence of the time-dependent Schroedinger
equations for most Hamiltonians that are relevant to known physics.

($\mathbf{ii}$) The other fundamental quantity of this formulation
is the generalized current density in field space defined as:

\begin{equation}
J_{k}(\{\varphi(\vec{x})\},t)\equiv P\frac{\delta S}{\delta\varphi_{k}(\vec{x})}\label{eq:(12)}\end{equation}

which strongly suggests the definition of the generalized local velocity
fields $\{\Phi(\vec{x},t)\}$ as solutions of the coupled functional
differential equations

\[
\Pi_{k}(\{\Phi(\vec{x},t)\},t\})\equiv\]

\begin{equation}
[\frac{\delta S}{\delta\varphi_{k}(\vec{x})}]_{\{\varphi(\vec{x})\}=\{\Phi(\vec{x},t)\}}\equiv[\frac{1}{P}Im\{\Psi^{*}\frac{\delta\Psi}{\delta\varphi_{k}(\vec{x},t)}\}]_{\{\varphi(\vec{x})\}=\{\Phi(\vec{x},t)\}}\label{eq:(13)}\end{equation}

This job is considerably simplified in practice by first Fourier transforming.

I shall consider eq(8), eq(9) and eq(13) as part of the set of fundamental
equations of the Bohm formulation of an ontological causal RQFT. However,
we use eq(13) in the following discussions, if needed.

I emphasize that Bohm-Hiley ontological reformulation of RQFT always
treats Bose fields as continuous distributions in spacetime - basically
because these quantum fields have perfectly well-defined classical
analogs. The textbook spin-0, spin-1 and spin-2 bosons, such as the
Higgs, photons, gluons, electroweak bosons and gravitons {[}18{]}
are, according to this viewpoint, not {}``particles'' in any naive
sense of this word, but just dynamical structural features of coupled
continuous scalar, vector and symmetric tensor fields, that first
become manifest when interactions with matter particles (elementary
or otherwise) occur {[}2{]},{[}8{]},{[}16{]}, as we shall illustrate.
This is obviously not just a question of semantics!

\subsection{Fourier transforms}

Practical work is greatly simplified if we Fourier transform first,
taking the advantages of the Poincaré invariance of relativistic field
theories. Fock space techniques {[}19{]} are more expedite then the
Schroedinger representation techniques in problems involving only
one or two field modes. Once the time-dependent Schroedinger wavefunction
is found, then we can immediately proceed to fix all the key quantities
of the BP.

The important field in this paper is the spin-1 Maxwell electromagnetic
field. The free field Hamiltonian is in a functional Schroedinger
representation {[}2{]},{[}19{]}

\begin{equation}
H_{0Maxwell}=\int d^{3}\vec{x}\;:[-\frac{\hbar^{2}}{c^{2}}\frac{\delta^{2}}{\delta\vec{\mathbf{A}}_{T}^{2}(\vec{x})}+\vec{\nabla}\times\vec{\mathbf{A}}_{T}(\vec{x}).\vec{\nabla}\times\vec{\mathbf{A}}_{T}(\vec{x})]:\label{eq:14}\end{equation}

where the symbol :: stands for {}``normal products'' {[}19{]}. The
classical vector field $\vec{\mathbf{A}}_{T}(\vec{x},t)$ is related
to the classical electric and magnetic field as follows:

\begin{equation}
\vec{\mathbf{E}}(\vec{x},t)=-\frac{1}{c}\frac{\partial}{\partial t}\vec{\mathbf{A}}_{T}(\vec{x},t)\qquad\vec{\mathbf{H}}(\vec{x},t)=\vec{\nabla}\times\vec{\mathbf{A}}_{T}(\vec{x},t)\label{eq:15}\end{equation}

We can transform to the Fock representation {[}19{]}

\begin{equation}
\vec{\mathbf{A}}_{T}(\vec{x})=\sum_{\lambda=\pm1}\int\frac{d^{3}\vec{k}}{\sqrt{(2\pi)^{3}2|\vec{k}|}}e^{i\vec{k}.\vec{x}}\vec{\mathbf{e}}(\vec{k},\lambda)a(\vec{k},\lambda)+h.c.\qquad\label{eq:16}\end{equation}

that must satisfy the basic commutation relations

\begin{equation}
[a(\vec{k},\lambda),a^{\dagger}(\vec{k}',\lambda')]=i\delta_{\lambda\lambda'}\delta(\vec{k}-\vec{k}')\qquad zero\; otherwise\label{eq:17}\end{equation}

and the conditions

\begin{equation}
\vec{\mathbf{e}}(\vec{k},\lambda).\vec{\mathbf{e}}(\vec{k'},\lambda')=\delta(\lambda,\lambda')\delta(\vec{k}-\vec{k}')\label{eq:18}\end{equation}

The conventions assumed in this paper are:

\begin{equation}
\vec{\mathbf{e}}(-\vec{k},+1)=-\vec{\mathbf{e}}(\vec{k},+1)\qquad\qquad\vec{\mathbf{e}}(-\vec{k},-1)=+\vec{\mathbf{e}}(\vec{k},-1)\label{eq:19}\end{equation}

\begin{equation}
\vec{\mathbf{e}}(\vec{k},\lambda).\vec{\mathbf{e}}(-\vec{k},\lambda')=(-1)^{(\lambda+1)/2}\delta(\lambda,\lambda')\qquad\label{eq:20}\end{equation}

and lead to thefinite de zero-th Maxwell Hamiltonian:

\begin{equation}
H_{0Maxwell}=\sum_{\lambda=\pm1}\int d^{3}\vec{k}\;|\vec{k}|\; a^{\dagger}(\vec{k},\lambda)a(\vec{k},\lambda)\qquad\qquad a(\vec{k},\lambda)|0>=0\label{eq:21}\end{equation}

\subsection{Many-body Dirac wavefunctions for point-like beable particles}

The relationships between fundamental classical relativity {[}2{]},{[}3{]},
and the Bohm formulation of non-relativistic Quantum Field theory
has a long history, going back to the fifties. There have been recent
new interesting proposals by B. Hiley and co-workers on this subject
{[}6{]},{[}7{]}, but in the present paper we follow the original ideas,
though closely linking them with contemporary textbook RQFT. Bohm
and co-workers went back to basics and did make an $a\; priori$ distinction
between $continuous$ beable bosonic fields (having natural classical
analogs) and $discrete$ elementary beable point-like particles, assumed
to be always guided by multi-dimensional Dirac wavefunctions. I should
stress at this point that the concept {}``beable'', as used in this
paper, has the same connotations as it had in the latest published
papers by Bohm and co-workers and also by J.S.Bell (who actually appears
to have coined it {[}3{]}). So present day Quantum Theory is assumed
in this paper to operate with basically 3 kinds of beables, $all$
equally necessary for the self-consistency of its mathematical formalism:
particles, fields and generally infinite dimensional time-dependent
Schroedinger wavefunctions.

Classical Mechanics does need only the first two beables. The third
kind of beables (wavefunctions) is unnecessary for its internal self-consistency
and completness; therefore it has no role to play in Classical Physics.
It is therefore essential to keep in mind that the BP formulation
is basically a $theory\; of\; beables.$ 

Forgetting this may lead to unnecessary confusion and endless discussions,
as earlier experience shows. The concept of beables is of course totally
foreign to textbook RQFT that deals solely with so-called {}``observables''
. According to the views underlying the present paper, all {}``observables''
are more or less complicated functions of beables, many specifically
taylored to the needs of the present-day experimental practices. The
inverse statement is not necessarily correct. I stress again, observables
in this sense are not fundamental to the BP (subsection 3.4).

It is assumed that the definition of the Dirac Hamiltonian includes
the minimal coupling prescription to all vector gauge fields (subsection
2.3). This could actually serve as an heuristic definition of what
one could mean with the words {}``elementary particles'', when applying
the BP to physically well-defined problems. For example, a {}``bare
electron'' is just a convenient mathematical abstraction in Quantum
Electrodynamics, because the dimensionless fine structure constant
is so small. However, it is not a gauge invariant object and so it
is - by definition - unphysical. Candidates to physical quantities
must be by definition gauge invariant objects. This implies among
other things that an electron at rest is always accompanied by its
Coulomb field in a Lorentz and gauge covariant manner : it always
appears and disappears together with its Coulomb field for all {}``observers''.

An obvious option from the ontological BP viewpoint (even if, according
to its authors, it might sound somewhat $ad\: hoc$ {[}2{]},{[}16{]})
is to consider leptons and quarks as present-day candidates to point-like
$beable$ {}``particles'', though characterized by their $spatial\; positions$
$only$ {[}1{]},{[}2{]},{[}3{]}. One should add and emphasize that
they are always supposed to be $actively$ guided by many-body beable
Dirac wavefunctions, obeying  the Dirac equation of motion and satisfying
the principle of minimal coupligs, say to $SU_{c3}\otimes SU_{2L}\otimes U_{1Y}$
vector gauge fields{[}2{]}. This assumption is perfectly consistent
with the remark that all so-called {}``measurements'' in experimental
physics ultimately boil down to {}``position measurements'' {[}2{]},{[}3{]}.

Let us be specific: assume an isolated Dirac particle having an inertial
mass $m_{0}$ and electric charge $-e_{0}$ :

\begin{equation}
[-i\vec{\alpha}.\vec{\nabla}_{\vec{x}}+\beta m_{0}]\psi(\vec{x})=E\psi(\vec{x})\label{eq:22}\end{equation}

\begin{equation}
[-i\vec{\alpha}.\vec{\nabla}_{\vec{x}}+\beta m_{0}]\bar{\psi}_{n}(\vec{x})=E\bar{\psi}(\vec{x})\qquad\bar{\psi}(\vec{x})\equiv\psi^{\dagger}(\vec{x})\beta\label{eq:23}\end{equation}

The equation of motion is the Dirac equation:

\begin{equation}
[-i\vec{\alpha}.\vec{\nabla}_{\vec{x}}+\beta m_{0}]\Psi(\vec{x},t)=i\hbar\frac{\partial}{\partial t}\Psi(\vec{x},t)\label{eq:24}\end{equation}

The (hermitean) Dirac matrices are given in the Dirac representation

\begin{equation}
\beta=\left(\begin{array}{cc}
1 & 0\\
0 & -1\end{array}\right)\qquad\vec{\alpha}=\left(\begin{array}{cc}
0 & \vec{\sigma}\\
\vec{\sigma} & 0\end{array}\right)\qquad\gamma_{5}=\left(\begin{array}{cc}
0 & 1\\
1 & 0\end{array}\right)\label{eq:25}\end{equation}

So, the additional equation of motion for this beable particle that
is necessary for $completing$ the list of quantum dynamical equations
is $postulated$ by the Bohm school to be {[}2{]},{[}16{]}:

\begin{equation}
\vec{v}(\{\vec{r}(t)\},t)=\{\frac{\Psi^{\dagger}(\vec{x},t)\times\vec{\alpha}\times\Psi(\vec{x},t))}{\Psi^{\dagger}(\vec{x},t)\times\Psi(\vec{x},t))}\}_{\vec{x}=\vec{r}(t)}\label{eq:26}\end{equation}

This replaces the guiding condition for unrelativistic particles,
but reduces to it in the non-relativistic limit. Not only is this
fully consistent with this non-relativistic limit $|\vec{v}|\ll1$
but also it is fully consistent with fundamental relativity {[}2,16{]},{[}17{]}.

Consider now a new physical situation in which the beable particle
finds itself in a quantum medium (i.e., the quantum vacuum) at least
partly represented by fields $\{\varphi(\vec{x})\}.$ Then the basic
ansatz (26) can be properly extended to apply to this more reallistic
situation, as we shall see. In this more reallistic case, the modified
system of equations (26) becomes a highly non-linear system of coupled
ordinary differential equations, just as in the non-relativistic many-body
case {[}1{]},{[}2{]}, and as such one should expect the appearence
of various kinds of singularities, when solving them {[}2{]}, {[}20{]}.
Numerical solutions show for example cases of bifurcation points even
in the simplest examples, which in our context could signal a quantum
transition. Also the transition to chaotic motions can be illustrated,
as the number of active degrees of motion increases even slightly
{[}2{]}.

A somewhat similar feature occurs of course also in classical newtonian
dynamics.

All of this plays a major role in the Bohm-Hiley ontological quantum
theory of measurements and quantum theory of transitions in general,
in which the archaic idea of a {}``collapse'' of the wavefunction
upon measurement is meaningless, and thus plays no role at all.

We emphasize that by now we are way beyond what is oficially declared
{}``meaningful''. This holds of course if one arbitrarily insists
in tacitly accepting the never clearly stated, let alone proved, implicit
metaphysics of standard textbook formulations (lucid deiscussions
about this can be found in ref {[}2{]}).

The basic reason is the same as in the non-relativistic many-body
case, i.e. the fundamental intrinsic quantum non-locality buried in
the basic definitions (13) and (26), which cannot be given any definite
meaning at all as far as textbooks are concerned.

So it is obvious that the velocities and quantum accelerations of
any particle are fundamentally $quantum$ correlated with the velocities
and accelerations of all particles in the Universe, regardless of
whether or not there are any classically describable potentials -
in stark contradiction with that we are accustomed to believe.

Let us emphasize again some of the central ideas as far as {}``beable
particles'' are concerned:

($\mathbf{i}$) the symbol $\{\vec{x}_{i}\}(i=1,2,3...)$ is reserved
for the set of eigenvalues of space coordinates operator in the Schroedinger
representation for particle wavefunctions, which one recalls is diagonal
in the spatial coordinates;

($\mathbf{ii}$) therefore, it must NOT be confused with the different
symbol $\{\vec{r}_{i}(t)\}$ for a collection of point-like {}``particle
spatial positions'', which are in general time-dependent and whose
ensemble averages (see the following section) have basically the same
connotations as those in Classical Physics (unprecisely speaking -
{}``observables''); note that no hidden semi-classical assumptions
are involved here at all.

($\mathbf{iii}$) a warning: in the Bohmean literature one often finds
the loose word {}``trajectory $\vec{r}$(t)'' to mean solutions
of the equations (26), given some initial and boundary conditions.
This might led unaware particle theoreticians to instinctively associate
this idea with their own familiar concept of {}`` Feynman paths''. 

However, the BP histories have nothing to do with the Feynman path
concept.

With this warning, I shall rather use the more neutral word particle
and field {}``histories'', instead of {}``trajectories''.

($\mathbf{iv}$) One must keep in mind that the infinite Dirac sea
of occupied negative frequency states $|0_{DIRAC}>$ does participate
in principle in every single step of any solution. This is of course
also true for textbooks. But fortunately this is isn't as bad as it
sounds, because of the separability, or factorization, of the total
wavefunction in an infinity of finite physical linked and unlinked
connected clusters {[}1{]},{[}2{]}. Thus one can verify that only
very few of these particles really participate at a time in any specific
case, and the rest simply drops out of sight in any meaningful and
doable experiments (this occurs obviously also in standard many-body
theory and RQFT, as it must of course). The famous Cluster Decomposition
Principle in Quantum Field Theory and in Many-Body theory shows how
and why this happens {[}17{]}, {[}18{]}. Precisely the same applies
to Boson QFT of subsection (2.1). 

Summing up, the set of coupled non-linear ordinary differential equations
for beable histories( (26) and its generalizations in section 3) together
with the similar equations (13) for Bose-Einstein fields, plus $all$
the initial and boundary condiditions on the wavefunction, constitutes
the $complete$ quantum dynamical framework that replaces the fundamental
laws of motion and interpretative schemes of Classical Mechanics and
Field Theory, relativistic or not. 

Detailed confrontation with the experimental data is $the$ worthwhile
task ahead. So, Its full range of validity and/or usefulness is ultimately
a matter that only  future theoretical $and$ experimental practices
can meaningfully decide.

To finish this section, let us consider an elementary illustrative
exercise: imagine a single Dirac free particle with mass $m_{0}$
guided by a 4-spinor plane wavefunction.

$\mathbf{\mathbf{(i)}\; A\; particle}\;\mathbf{guided}\;\mathbf{by\; a\;}\mathbf{positive\; frequency\; wavefunction}$

We begin by defining a standard reference system, say K. Let a particle
of species n and rest mass $m_{0n}$ move along K's positive z-axis
$\vec{\mathbf{e}}_{3}$ guided by the positive energy Dirac plane
wave

\begin{equation}
\vec{p}=p\:\mathit{\vec{\mathbf{e}}_{3}}\qquad\qquad E_{n}(p)=+\sqrt{p^{2}+m_{0n}^{2}}\label{eq:27}\end{equation}

Let the $\chi_{\lambda}(\vec{\mathbf{e}}_{3})$ be an helicity eigenstate,
part of the positive frequency wavefuncion $w_{n\vec{p}\lambda}^{(+)}(\vec{p})$:

\begin{equation}
\mathbf{\vec{\mathbf{\boldsymbol{\sigma}}}}.\vec{p}\chi_{\lambda}(\vec{\mathbf{e}}_{3})=2p\lambda\chi_{\lambda}(\vec{\mathbf{e}}_{3})\qquad\lambda=\pm\frac{1}{2}\label{eq:28}\end{equation}

\begin{equation}
w_{n\vec{p}\lambda}^{(+)}(\vec{p})\equiv\sqrt{\frac{E_{n}(p)+m_{0n}}{2E_{n}(p)}}\left(\begin{array}{c}
1\\
\frac{\mathbf{\vec{\mathbf{\boldsymbol{\sigma}}}}.\vec{p}}{E_{n}(p)+m_{0n}}\end{array}\right)\chi_{\lambda}(\vec{e}_{3})\times\varphi_{n}\label{eq:29}\end{equation}

\begin{equation}
\psi_{n\vec{p}\lambda}^{(+)}(x,y,z,t)=w_{n\vec{p}\lambda}^{(+)}(\vec{p})e^{-iE_{n}(p)t+i\vec{p}.\vec{x}}\label{eq:30}\end{equation}

The {}``internal'' flavour-colour-... component of the wavefunction
is called collectively $\varphi_{n}$ and is supposed to be an irreducible
representation of the Lie algebra $SU_{c3}\otimes SU_{2L}\otimes U_{1Y}.$ 

Let us calculate the velocity $\vec{v}(\vec{r}(t$)) at the position
$\vec{r}(t)$ of a beable particle under guidance of this 4-spinor
wavefunction: 

\[
\psi^{\dagger}(\vec{x},t)\times\vec{\alpha}\times\psi(\vec{x},t)=\]

\begin{equation}
=\frac{E(p)+m_{0n}}{2E(p)}\varphi_{n}^{\dagger}\chi_{\lambda}^{\dagger}(\vec{e}_{3})\left(\begin{array}{cc}
1 & \frac{\mathbf{\vec{\mathbf{\boldsymbol{\sigma}}}}.\vec{p}}{E(p)+m_{0}}\end{array}\right)\vec{\alpha}\left(\begin{array}{c}
1\\
\frac{\mathbf{\vec{\mathbf{\boldsymbol{\sigma}}}}.\vec{p}}{E(p)+m_{0}}\end{array}\right)\chi_{\lambda}(\vec{e}_{3})\times\varphi_{n}\label{eq:31}\end{equation}

\[
\psi^{\dagger}(\vec{x},t)\psi(\vec{x},t)=\]

\begin{equation}
=\frac{E(p)+m_{0}}{2E(p)}\varphi_{n}^{\dagger}\chi_{\lambda}^{\dagger}(\vec{e}_{3})\left(\begin{array}{cc}
1 & \frac{\mathbf{\vec{\mathbf{\boldsymbol{\sigma}}}}.\vec{p}}{E(p)+m_{0}}\end{array}\right)\left(\begin{array}{c}
1\\
\frac{\mathbf{\vec{\mathbf{\boldsymbol{\sigma}}}}.\vec{p}}{E(p)+m_{0}}\end{array}\right)\chi_{\lambda}(\vec{e}_{3})\times\varphi_{n}\label{eq:32}\end{equation}

So the beable particle velocity when it finds itself at the locality
$\vec{r}(t)$ in some inertial reference frame is given by

\begin{equation}
\vec{v}(\{\vec{r}(t)\},t).\vec{e}_{3}=\frac{1}{E(p)}\vec{p}.\vec{e}_{3}\label{eq:33}\end{equation}

It moves parallel to the z-axis at the constant velocity $\vec{v}$,
as expected.

$\mathbf{\mathbf{(ii)}\; A\; particle}\;\mathbf{guided}\;\mathbf{by\; a\;}\mathbf{negative\; frequency\; wavefunction}$

Similarly, consider a beable particle guided by the negative frequency
Dirac wavefunction:

\begin{equation}
\psi_{n\vec{p}\lambda}^{(-)}(\vec{x},t)=w_{n\vec{p}\lambda}^{(-)}(\vec{p})e^{iE_{n}(p)t+i\vec{p}.\vec{x}}\label{eq:34}\end{equation}

\begin{equation}
w_{n\vec{p}\lambda}^{(-)}(\vec{x},t)=\sqrt{\frac{E_{n}(p)-m_{0n}}{2E_{n}(p)}}\left(\begin{array}{c}
1\\
-\frac{\vec{\mathbf{\boldsymbol{\sigma}}}.\vec{p}}{E_{n}(p)-m_{0n}}\end{array}\right)\chi_{\lambda}(\vec{e_{3}})\times\varphi_{n}\label{eq:35}\end{equation}

Its velocity is then

\begin{equation}
\vec{v}(\vec{r}(t),t)=\frac{\chi_{\lambda}^{\dagger}(\vec{e_{3}})\left(\begin{array}{cc}
1 & -\frac{\vec{\mathbf{\boldsymbol{\sigma}}}.\vec{p}}{E(p)-m_{0}}\end{array}\right)\vec{\alpha}\left(\begin{array}{c}
1\\
-\frac{\vec{\mathbf{\boldsymbol{\sigma}}}.\vec{p}}{E_{n}(p)-m_{0n}}\end{array}\right)\chi_{\lambda}(\vec{e_{3}})}{\chi_{\lambda}^{\dagger}(\vec{e_{3}})\left(\begin{array}{cc}
1 & -\frac{\vec{\mathbf{\boldsymbol{\sigma}}}.\vec{p}}{E(p)-m_{0}}\end{array}\right)\left(\begin{array}{c}
1\\
-\frac{\vec{\mathbf{\boldsymbol{\sigma}}}.\vec{p}}{E_{n}(p)-m_{0n}}\end{array}\right)\chi_{\lambda}(\vec{e_{3}})}\label{eq:36}\end{equation}

which gives

\begin{equation}
\vec{v}(\vec{r}(t),t).\vec{e}_{3}=-\frac{\vec{p}.\vec{e}_{3}}{E(p)}\label{eq:37}\end{equation}

and the beable particle moves antiparallel to the z-axis,i.e. with
the velocity $-\vec{v}$, as expected.

In modern standard textbooks, this would be considered to be a rather
odd and archaic way of talking, although there is nothing wrong with
it; it is about this stage that one shifts to the more practical abstract
field theoretic (or many-body) descriptions, by introducing the concept
of {}``antiparticles'' (or {}``holes'' in the Fermi sea of many-body
fermion systems). This is because this is more practical, if one follows
the standard way of thinking and practicing. Moreover, it is more
convenient to talk in this way when directly dealing with the experimental
data.

It is none the less more convenient for us, given the entirely different
conceptual basis of the Bohm-Hiley causal interpretation, to keep
to (13) and (26) {[}2{]}.

Finally: given some initial conditions on the wavefunction, what is
the probability that a definite particle spatial distribution $\{\vec{r}_{n}(t)$\}
and velocity distribution $\{\vec{v}_{n}(t)$\} , together with a
field distribution $\{\Phi(\vec{x},t)\}$ occurs at any time t in
some inertial reference frame? The answer is:

\begin{equation}
P(\{\vec{r}_{n}(t)\},\{\Phi(\vec{x},t)\};t)=|\Psi(\{\vec{r}_{n}(t)\},\{\Phi(\vec{x},t)\});t)|^{2}\label{eq:38}\end{equation}
$\{\vec{r}_{n}(t)\}$ ($\{\Phi(\vec{x},t)\}$) are particle (field)
histories in this vacuum that satisfy all the relevant initial conditions
on the wavefunction {[}1{]},{[}2{]}.

\subsection{Ensemble averages and connections to both standard RQFT and experimental
physics}

We emphasize that the BP interprets Quantum Theory as a causal metatheory
of beables. Therefore, the countless counterarguments against it that
can be found dispersed in the literature of the last 30 years or so
are in most cases (to my mind at least) besides the point.

Nevertheless, it is quite legitimate to ask - what has the BP of RQFT
to do with the standard textbook accounts and the present day experimental
practices?

This was clearly thoroughly answered and explained by Bohm not only
in his very first Physical Review papers on this subject in the early
fifties but also in all his subsequent published work {[}2{]}. Thus
only the shortest possible summary is given here.

One begins with the manifold of solutions of the above dynamical equations
of motion, satisfying the appropriate boundary and initial conditions
on the wavefunction. Let us next consider the desired connection to
the textbooks and to the present conditions of experimental research.
The suggested procedure is that one should proceed to apply standard
schostatic methods and compute statistical averages EA over pure ensembles
of such histories (i.e. over all inacessible data on initial positions
and velocities of particles and field configurations).

As an example, let us simplify the case to just one spatial dimension
without losing generality:

\begin{equation}
<X(t)>_{EA}\equiv\int dX(t)\Psi^{\dagger}(X(t),t)\; X(t)\;\Psi(X(t),t)\label{eq:39}\end{equation}

\begin{equation}
<P(t)>_{EA}\equiv\int dX(t)\Psi^{\dagger}(X(t),t)\:[-i\frac{\partial\Psi(X(t),t)}{\partial X(t)}]\label{eq:40}\end{equation}

\begin{equation}
<E(t)>_{EA}\equiv\int dX(t)\Psi^{\dagger}(X(t),t)\:[i\frac{\partial\Psi(X(t),t)}{\partial t}]\label{eq:41}\end{equation}

In the general case $\Psi(t)$ is made of sums over products of Dirac
4-spinors and the Boson fields represented in the wavefunction, that
satisfy the given boundary and initial conditions. These expectation
values are directly related to standard textbook prescriptions and
therefore to the experimental data. Further discussions can be found
in {[}1{]} and {[}2{]}.

\section{Case Studies}

We are now ready to briefly discuss a few illustrative cases that
played, and still do play, important roles in the overall development
of contemporary Particle Physics.

\subsection{Vacuum survival probabilities of a single positronium atom at rest}

To problem of understanding and explaining how any atom could be stable
(and thus exist at all) ignated the quantum revolution of the XXth
century. The reason is of course known to every undergraduate to-day:
it was simply that inevitable and fundamental classical predictions
were in blatant contradiction with reality. Based on this fact, one
can imagine an $a\; posteriori$ philosophizing as follows:

(i) it was fortunate that by 1913 one already had a pretty good experimental
description of the behaviour of the simplest of all existing atoms
- the H-atom;

(ii) it was fortunate that in 1913 no one knew about positronium atoms
{[}21{]}.

Let us see how the BP would $explain$ such obvious contradictions
among our {}``observables'' and our most fundamental pre-XXth century
preconceptions.

Let us imagine a single isolated positronium atom at rest somewhere
in the Universe. Any isolated positronium atom, even in its $ground$
state, is unstable! This is not the case, however, with the H-atom,
simply because of exact conservation laws, plus the still ultimately
unexplained fact that a d-quark is heavier then a u-quark.

Let us look in particular to a Positronium (Ps) atom where there are
no extra complications due to the strong interaction : we begin with
a model Hamiltonian

\begin{equation}
H_{eff}=U_{0}+H_{e^{+}e^{-}}+H_{0Maxwell}+V_{0e^{+}e^{-}}\label{eq:42}\end{equation}

where the first term on the rhs $U_{0}$ is an arbitrary constant,
which can be trivially renormalized away.

The use of natural units $\hbar=1=c$ will be temporarily suspended
in this subsection.

The next term in definition (42) is the Ps Hamiltonian (standard notation
e.g. {[}20{]}):

\begin{equation}
H_{e^{+}e^{-}}=-\frac{\hbar^{2}\nabla_{X}^{2}}{4m_{0}}-\frac{\hbar^{2}}{m_{0}}\nabla_{\rho}^{2}-\frac{\alpha}{\rho}\label{eq:43}\end{equation}

\begin{equation}
m_{0}=electron\;(positron)\; rest\; mass\label{eq:44}\end{equation}

\begin{equation}
\rho=|\vec{x}_{e^{-}}-\vec{x}_{e^{+}}|\label{eq:45}\end{equation}

The CM of the atom is assigned the arbitrary position

\begin{equation}
\vec{X}\equiv\frac{\vec{x}_{e^{-}}+\vec{x}_{e^{+}}}{2}\label{eq:46}\end{equation}

We consider only atomic bound states: introduce

\begin{equation}
H_{0e^{+}e^{-}}\equiv H_{e^{+}e^{-}}-\frac{|\vec{P}|^{2}}{4m_{0}}\label{eq:47}\end{equation}

\begin{equation}
H_{0e^{+}e^{-}}\psi_{\nu}(\vec{\rho})=\varepsilon_{\nu}\psi_{\nu}(\vec{\rho})\label{eq:48}\end{equation}

\begin{equation}
\varepsilon_{\nu}\Rightarrow-|\varepsilon_{nL}|\label{eq:49}\end{equation}

The ground state wavefunctions n=1 L=0 ($\equiv$1S) are degenerate
in the stated approximation (i.e. spin independent Hamiltonian).

A good quantum number that can further label these states is charge
conjugation C:

\begin{equation}
C=(-1)^{L+S}=(-1)^{S}=\pm1\label{eq:50}\end{equation}

Thus the spin singlet (C=+1) and spin triplet (C=-1) are both 1S ground
states in this approximation, but even small spin-dependent residual
forces can lift this degeneracy. This possibility is left out in the
following discussions.

The electromagnetic field is represented by the Hamiltonian in momentum
space:

\begin{equation}
H_{0Maxwell}=c\sum_{\lambda=\pm1}\int d^{3}\vec{p}\;|\vec{p}|\; a^{\dagger}(\vec{p},\lambda)a(\vec{p},\lambda)\label{eq:51}\end{equation}

\begin{equation}
a(\vec{p},\lambda)|0>=0\quad a^{\dagger}(\vec{p},\lambda)|0>\equiv|\vec{p},\lambda>\label{eq:52}\end{equation}

\begin{equation}
\frac{1}{\sqrt{1+\delta(\vec{p}_{1}\lambda_{1},\vec{p}_{2}\lambda_{2})}}a^{\dagger}(\vec{p}_{2},\lambda_{2})a^{\dagger}(\vec{p}_{1},\lambda_{1})|0>\equiv|\vec{p}_{2},\lambda_{2};\vec{p}_{1},\lambda_{1}>\label{eq:53}\end{equation}

and so forth. The effective interaction is defined as

\begin{equation}
V_{0e^{+}e^{-}}=e_{0}V_{0e^{+}e^{-}}^{(1)}+e_{0}^{2}V_{0e^{+}e^{-}}^{(2)}\label{eq:54}\end{equation}

and can be put in the form

\begin{equation}
e_{0}V_{0e^{+}e^{-}}^{(1)}=\sum_{\lambda=\pm1}\int\frac{d^{3}\vec{k}}{\sqrt{(2\pi)^{3}2|\vec{k}|}}F_{\lambda}(\vec{k},\vec{\rho})[a(\vec{k},\lambda)+a^{\dagger}(\vec{k},\lambda)]\label{eq:55}\end{equation}

\begin{equation}
F_{\lambda}(\vec{k},\vec{\rho})=\frac{\hbar e_{0}}{2m_{0}c}\; sin(\frac{\vec{k}.\vec{\rho}}{2})\;\mathbf{\vec{k}}.\vec{\mathbf{e}}(\vec{k},\lambda)\label{eq:56}\end{equation}

and

\begin{equation}
e_{0}^{2}V_{0e^{+}e^{-}}^{(2)}=\frac{e_{0}^{2}}{2m_{0}c^{2}}\int d^{3}\vec{x}\;:\mathbf{\vec{A}}_{T}(\vec{x}).\vec{\mathbf{A}}_{T}(\vec{x}):\label{eq:57}\end{equation}

The interaction piece (57) turns out to be spurious in our context,
and will henceforth be dropped.

A non-relativistic Hamiltonian is a great formal simplification, but
it is of course not an essential assumption for the purposes of this
paper. The coupling to the electromagnetic field results from the
standard principle of a minimal coupling {[}18{]},{[}19{]},{[}21{]}.
Conservation laws then decide which of the two above mentioned degenerate
states, if any, is stable against annihilation into at least 2 gammas
(C=+1). The same conservation laws show that the other state is also
unstable, but decaying predominantly into 3 gammas (C=-1). These so-called
open channels may contribute to the S-matrix, of course.

As mentioned, the first step for implementing our program is to find
the guiding beable time-dependent Schroedinger wavefunction:

\begin{equation}
[H_{e^{+}e^{-}}+H_{0Maxwell}+e_{0}V_{0e^{+}e^{-}}^{(1)}]|\Psi(t)>=i\hbar\frac{\partial}{\partial t}|\Psi(t)>\label{eq:58}\end{equation}

The Schroedinger time-dependent wavefunctional is defined to be 

\begin{equation}
\Psi(\vec{\rho},\vec{\mathbf{A}}_{T}(\vec{x});t)\equiv<\vec{\rho;}\vec{\mathbf{A}}_{T}(\vec{x})|\Psi(t)>\label{eq:59}\end{equation}

We are interested in the physical vacuum expectation value of this
wavefunctional.

Unfortunately, a real understanding of the physical vacuum is quite
beyond our present capabilities. One has then to proceed as usual,
by adapting similar methods known to be reliable, e.g. many-body ground
state methods - and hope that they somehow work also in the present
context, at least up to some point! So we $assume$ that

\begin{equation}
<0_{DIRAC};0|0_{phys}>\neq0\label{eq:60}\end{equation}

\begin{equation}
|0_{phys}>\longleftrightarrow|0_{DIRAC};0><0_{DIRAC};0|0_{phys}>+...\label{eq:61}\end{equation}
 $|0>$is taken to represent the ground state of an infinite collection
of non-interacting 3-D harmonic oscillators, whereas $|0_{DIRAC}>$represents
the Dirac fermion vacuum. Explicit expressions for the perturbative
wavefunction in the rest system to any order can be found e.g. in
ref {[}21{]}:

\begin{equation}
\Psi(\vec{\rho};\vec{\mathbf{A}}_{T}(\vec{x});t)\equiv<\vec{\rho};\vec{\mathbf{A}}_{T}(\vec{x})|\Psi(t)>\label{eq:62}\end{equation}

\begin{equation}
|\Psi(t)>=\sum_{N=0}^{\infty}\sum_{\nu}\sum_{\{n\}}C_{\nu\{n\}}^{(N)}(t)\; exp[i|\varepsilon_{\nu}|t/\hbar]\:|\nu>\times exp\{-iE(\{n\}t/\hbar)|\{n\}>\label{eq:63}\end{equation}

Plugging this definition into the Schroedinger equation of motion
(58) one finds a recursion formula for the coefficients $C_{\nu\{n\}}^{(N)}(t)$
for $t\geq0$:

\begin{equation}
C_{\nu\{0\}}^{(0)}(t)=C_{\nu\{0\}}(0)\qquad\mathbf{N=0}\label{eq:64}\end{equation}

and

\[
C_{\nu'\{n'\}}^{(N)}(t)=\frac{-i}{\hbar}\sum_{\nu}\sum_{\{n\}}<\nu';\{n'\}|V_{0e^{+}e^{-}}^{(1)}|\nu;\{n\}>\times\]

\begin{equation}
\times\int_{0}^{t}dt_{1}C_{\nu\{n\}}^{(N-1)}(t_{1})\times exp[-i(\omega(\nu',E\{n'\})t_{1}]\times exp[i(\omega(\nu,\{n\})t_{1}]\qquad\mathbf{N\geq1}\label{eq:65}\end{equation}

with the definition

\begin{equation}
\hbar\omega(\alpha,\{m\})\equiv-|\varepsilon_{\alpha}|+E\{m\}\label{eq:66}\end{equation}

and in general

\[
<\nu';\{n'\}|V_{0e^{+}e^{-}}^{(1)}|\nu;\{n\}>=\sum_{\lambda''=\pm1}\int\frac{d^{3}\vec{k}''}{\sqrt{(2\pi)^{3}2|\vec{k}''|}}I_{\nu'\nu}(\vec{k}'',\lambda'')\times\]

\begin{equation}
\times<\{n'\}|[a(\vec{k}'',\lambda'')+a^{\dagger}(\vec{k}'',\lambda'')]|\{n\}>\label{eq:67}\end{equation}

with the definition

\begin{equation}
I_{\nu'\nu}(\vec{k}'',\lambda'')\equiv\int d^{3}\vec{\rho}\psi_{\nu'}^{\dagger}(\vec{\rho})F_{\lambda''}(\vec{k}'',\vec{\rho})\psi_{\nu}(\vec{\rho})\label{eq:68}\end{equation}

Let us assume that at $t=0$ the state is $|I;0>$. Let us further
assume that there is some probability that at some later time t it
still is |I;0>. We would like to find that probability. Then our master
formula (65) leads to a prescription :

\begin{equation}
C_{I\{0\}}^{(0)}(t)=c_{I}\label{eq:69}\end{equation}

\begin{equation}
C_{I\{0\}}^{(1)}(t)\equiv0\label{eq:70}\end{equation}

\begin{equation}
C_{I\{0\}}^{(2)}(t)=(\frac{-i}{\hbar})^{2}\sum_{\nu,\{n\}\neq I,\{0\}}J_{\nu\{n\}}(t)\:|<\nu;\{n\}|V_{0e^{+}e^{-}}^{(1)}|I;\{0\}>|^{2}\times c_{I}\label{eq:71}\end{equation}

\begin{equation}
J_{\nu\{n\}}(t)\equiv\int_{0}^{t}dt_{1}\int_{0}^{t_{1}}dt_{2}exp[-i(\omega(I,\{0\})-\omega(\nu,\{n\}))(t_{2}-t_{1})]\label{eq:72}\end{equation}

Thus up to 2nd order the mode population in the initial and final
must differ by one vacuum mode.

The vacuum averaged desired Ps wavefunction then becomes

\begin{equation}
<\Phi(\vec{\rho};\vec{\mathbf{A}}_{T}(\vec{x});t)>\equiv\psi_{I}(\vec{\rho},t)=\{c_{I}+C_{I\{0\}}^{(2)}(t)+...\}\psi_{I}(\vec{\rho},0)\qquad t\geq0\label{eq:73}\end{equation}
 One can then proceed to the desired result, which is to find the
survival probability of some initial state $\psi_{I}(\vec{\rho},0)$
of the atom in its rest system in vacuum. The remaining of this calculation,
and its final outcome, can be found in ref {[}21{]}, but that is hardly
the point here. We are trying to:

(i) sketch how the BP explains the decay of an isolated Ps atom in
its ground state;

(ii) find out precisely what the spacetime dependence is, as we shall
need that in subsection (3.3).

Once the solution for the time-dependent wavefunction is found that
satisfies the initial conditions, then the next question would be:
what happens to an isolated beable {}``particle'' (i.e. a Ps atom)
in vacuum in its rest system if guided by this wavefunction (73) as
time goes by, starting from some well-defined initial state, say at
$t=0$? 

More precisely, one is asking to predict the precise time dependence
of the internal vector, given definite initial conditions:

\begin{equation}
\vec{r}(t)=\vec{r}_{e^{-}}(t)-\vec{r}_{e^{+}}(t)\label{eq:74}\end{equation}

The answer to the above question is given by solving the non-relativistic
edition of equation (26) {[}1{]},{[}2{]} with the definition (73)
for the guiding wavefunction when t$\geq0$. So the answer is:

\begin{equation}
\psi_{I}(\vec{\rho,}t)=\sum_{N=0}^{\infty}\psi_{I}^{(N)}(\vec{\rho,}t)\label{eq:75}\end{equation}

\begin{equation}
\vec{v}(\vec{r}(t),t)=\frac{1}{|\psi_{I}(\vec{r}(t),t)|^{2}}\{Im[\sum_{p=0}^{\infty}\sum_{m=0}^{p}\psi_{I}^{(p)}(\vec{\rho},t)\vec{\nabla}\psi_{I}^{(p-m)}(\vec{\rho},t)]\}_{\vec{\rho}=\vec{r}(t)}\label{eq:76}\end{equation}

Appropriate initial conditions have to be specified of course {[}1{]},{[}2{]}
before one can find the proper solution and (if possible) confront
it with textbook material and/or experimental clues).

One must not forget that conservation of probability (mandatory both
in SP and the BP) is fundamental and demands that at any arbitrary
time t :

\begin{equation}
|\psi_{I}(\vec{\rho},t)|^{2}=|\psi_{I}(\vec{\rho},0)|^{2}\label{eq:77}\end{equation}

The probability that the Ps atom that was in a specified initial state
(I) has survived at any time t>0 with the constituents in the relative
position $\vec{r}(t)$ $and$ rate of change $\vec{v}(\vec{r}(t),t)$
(81) is then

\begin{equation}
P_{I}(\vec{r}(t),t)=|\psi_{I}(\vec{r}(t),t)|^{2}\label{eq:78}\end{equation}

where $\vec{r}(t)$ is a proper solution of eq (76).

Note that we thus obtain a non-relativistic approximation for the
rate of change with time of the internal relative position vector
of the atomic constituents, as the atom stays at rest somewhere, in
agreement with our simplifying assumption about the non-relativistic
motions inside the atom.

If one can imagine that the vacuum coupling is switched-off, then
one would find that all Ps states would become fully stable {[}1{]},{[}2{]},
i.e. the internal relative separation between the constituents would
either not change with time for non-degenerate states, or change periodically
with time if degenerate, because in this latter case one could always
make linear combinations that could build up complex wavefunctions
{[}1{]},{[}2{]}.

This $description$ is $explained$ by the Bohm school by the balance
established between the classical attractive Coulomb force acting
between the $e^{\pm}$ constituent particles and the centrifugal quantum
forces associated to the guiding wavefunction, and originating from
the quantum active information potential {[}1{]},{[}2{]}. This replaces
the classical kinetic energy of the system $e^{\pm}$ that played
the same role in the old semi-classical Bohr atomic model. The basic
physics involved in this case was discussed in ref {[}1{]},{[}2{]}
and there is hardly any point in repeating that here.

But again: precisely what happens to any sufficiently isolated individual
Ps atom, if the coupling to the quantum medium is non-existent? The
BP does have a very concrete, precise and in principle checkable proposal,
applicable for $all$ times. 

Whether this proposal can, or cannot, be somehow be verified in some
distant future, perhaps using now unknowable experimental capabilities,
is another matter altogether. 

The situation is expected to change of course, if the Ps can only
exist in the universal (but mostly unknown) quantum medium, being
referred to here as the {}``vacuum''. Then, the rate with which
the beable relative position vector $\vec{r}$(t) changes with time
is still given by eq. (76). If all conservation laws are obeyed, especially
for energy and momentum, then the pertinent S-matrix elements (asymptotic
times) can be recovered e.g. for the relevant 2 and 3 $\gamma$-decays:
the original Ps atom thus becomes {}``metamorphosed'' into $\gamma$
radiation. This is just an example of a quantum transition.

These conclusions (except of course for the Ps atom internal velocity
$and$ position vectors) formally agree with the textbooks, but the
interpretations there (if given at all) are both totally different
{[}2{]},{[}3{]} and indeed inextrincably linked to the Quantum Theory
of Measurements. An essential difference is that in the BP this link
is not placed at the very theoretical core of the theory, and thus
has a completely different (and natural!) role to play (subsection
2.4). 

In most practical cases, however, one is only concerned only with
asymptotic times, even if in principle the theory can provide detailed
information on (usually very complicated) particle histories, for
all times, given of course the initial conditions. So far the rule
has been that there is full numerical agreement at asymptotic times
(i.e. in the S-matrix domain) with standard textbook RQFT results
and predictions.

\subsection{Neutrino/antineutrino flavour metamorphosis in vacuum}

Suppose that in some distant supernova explosion a neutrino is released
in some definite flavour state, say $\beta$. Let us further assume
that it then moves in free space towards the Earth, which it reaches
at local time T. If the distance to this supernova site is say L,
then$L$$\approx T$ in the relativistic units used in this paper.
This is because the neutrino moves in vacuum with almost the speed
of light c=1.

Let us now play with the following thought: imagine that an intelligent
undergraduate would like to have a genuine explanation of what $really$
goes on behind the curtain of phenomena of neutrino flavour oscillations
in vacuum (or within ordinary matter). He is now asked to make a short
list of questions to which he would expect to obtain real answers:

What happened to this free $\beta-$neutrino during the time interval
t=0 and t=T between the supernova explosion that gave birth to it
and its arrival to Earth?

How and why can there be flavour oscillating metamorphosis as the
neutrino moves along some trajectory, even if the neutrino first emerged
through a weak interaction process, thus in a definite flavour state
$\beta$?

Why, and precisely how, are these remarkable oscillations related
to the neutrino masses?

If a neutrino is a particle with a definite inertial mass, how come
we can {}``see'' flavour oscillations, which is typically a wave-like
phenomenon?

And so forth. The answers to these and many other similar questions
can in principle at least be both quantitatively and qualitatively
deduced and $explained$ by the Bohm formulation of Quantum Field
Theory. Whether the {}``explanation'' is {}``right'' or {}``wrong''
is a different matter altogether. Only further predictions and difficult
experimentation can decide that. We are discussing here only questions
of principle.

The experimental data on neutrinos strongly suggest that their masses
are very small compared to all other known leptons and quarks. We
are thus dealing with highly relativistic particles. 

The known neutrinos propagating in vacuum are supposed to interact
very weakly with the perennial virtual heavy electroweak vaccum fluctuating
modes $W^{\pm}$(section 2.2). The primary interaction causing these
electroweak spin flips is given by the Standard Model {[}18{]}:

\begin{equation}
l_{\beta\uparrow}\rightleftharpoons l_{\alpha\downarrow}+W^{+}\qquad\qquad l_{\beta\downarrow}\rightleftharpoons l_{\alpha\uparrow}+W^{-}\qquad\qquad\alpha,\beta=e,\mu,\tau\label{eq:79}\end{equation}

with the definitions

\[
1st\; lepton\; family:\qquad l_{e\uparrow}\equiv\nu_{e}\equiv electron\: neutrino;\qquad l_{e\downarrow}\equiv e\equiv electron;\]

and likewise for the 2nd family (muon) and the 3rd family (tau).

The BP of Quantum Mechanics agrees with the textbook formulations
that these {}``quantum medium'' couplings must be the primary ones
and the ultimate cause of the observed and famous flavour vacuum fluctuations
of neutrinos.

A qualitative and quantitative natural quantum theoretic explanation
of what goes on here, according to the Bohm school, would start by
writing down the beable neutrino guiding wavefunction in vacuum. As
a neutrino of any species n is a particle, it is natural that its
guiding wavefunction should be given by a wavepacket, say:

\begin{equation}
\Psi_{n\alpha}(\vec{x},t)=\psi_{n}(\vec{x},t)\times\varphi_{\alpha}\label{eq:80}\end{equation}

where the ansatz for the wavefunction is

\begin{equation}
\psi(\vec{x},t)=\sum_{n=1}^{3}\sum_{\nu=+,-}\int d^{3}\vec{p}C_{n}^{(\nu)}(\vec{p},t)\: w_{n}^{(\nu)}(\vec{p,}\omega_{n}^{(\nu)}(p))e^{-i\nu E_{n}t+i\vec{p.}\vec{x}}\qquad t\geq0\label{eq:81}\end{equation}

where

\begin{equation}
(\vec{\alpha}(n).\vec{p}+\beta(n)m_{0}(n))w_{n}^{(\nu)}(\vec{p},E_{n}(p))=\nu E_{n}(p)w_{n}^{(\nu)}(\vec{p},E_{n}(p))\label{eq:82}\end{equation}

\begin{equation}
E_{n}(p)\equiv+\sqrt{p^{2}+m_{0}^{2}(n)}\label{eq:83}\end{equation}

We have also a boundary and initial condition on the wavefunction

\begin{equation}
\Psi(\vec{x}=\vec{L},t=0)=F_{\beta}(L)\varphi_{\beta}\qquad t=0\label{eq:84}\end{equation}

The coefficients $C_{n}^{(\nu)}(\vec{p},t)$ in eq(85) are determined
by solving the Dirac equation of motion including the initial condition:

\begin{equation}
[-i\vec{\alpha}(n).\vec{\nabla}+\beta(n)m_{0}(n)+\sum_{\beta}V_{0S}(\alpha,\beta,W^{\pm})]\Psi_{n\alpha}(\vec{x},t)\}=i\frac{\partial}{\partial t}\Psi_{n\alpha}(\vec{x},t)\label{eq:85}\end{equation}

We assume that the dependence on the vacuum comes from the primary
and the quantum medium (=vacuum, quantum ether,...) induced effective
interaction $V_{0S}(\mu,\nu,W^{\pm})$.

This is not only suggested by the Standard Model {[}18{]} but to some
degree also calculable. The primary couplings are between flavour
lepton electroweak isodoublets $(\nu_{e},e),(\nu_{\mu},\mu),(\nu_{\tau},\tau)$
and the very heavy $W^{\pm}$ vacuum modes (components of an electroweak
isovector) and symbolized by the quantum transitions (79). We end
up with

\begin{equation}
\Psi_{\beta}(\vec{x},t)=\sum_{\alpha=e,\mu,\tau}\Phi_{\beta\alpha}(\vec{x},t)\varphi_{\alpha}\qquad\qquad t\geq0\label{eq:86}\end{equation}

where the coefficients $\Phi_{\beta\alpha}(\vec{x},t)$ are complicated
spacetime functions (i.e. space-,spin-,momentum-, energy-dependent)
of the neutrino masses to be obtained by solving the specific equation
(85). 

Note that in particular the neutrino spin plays only a very modest
role in all this. The paramount degrees of freedom involved here are
flavour-mass and flavour flips. We are interested on the time-dependence
of the wavefunction$\Psi_{\beta}(\vec{x},t)$.

It can be shown {[}22{]},{[}23{]} that the kinematical conditions
here relevant are such that they allow one to make the reasonably
good estimate

\begin{equation}
\Phi_{\beta\alpha}(|\vec{x}|\approx L,t\approx L)=\sum_{n=1}^{3}[\mathbf{U}_{\nu}]_{\alpha n}\: e^{-im_{n}^{2}L/2\overline{E}}\:[\mathbf{U}_{\nu}^{\dagger}]_{n\beta}\label{eq:87}\end{equation}

\begin{equation}
\mathbf{U}_{\nu}^{-1}=\mathbf{U^{\dagger}}_{\nu}\label{eq:88}\end{equation}

where $\overline{E}$ is a properly defined neutrino mean energy.
The unitary matrix $\mathbf{U}_{\nu}$ relates by definition the $flavour\: eigenstates$
$\varphi_{\alpha}$ to the $mass\; eigenstates$ $\chi_{k}$ :

\begin{equation}
\varphi_{\alpha}=\sum_{k=1}^{3}[\mathbf{U}_{\nu}]_{\alpha k}\chi_{k}\qquad\alpha=\nu_{e},\nu_{\mu},\nu_{\tau}\label{eq:89}\end{equation}

Most of our vital experimental clues are encoded in this neutrino
{}``mapping-matrix'' $\mathbf{U}_{\nu}$ (and its hadronic equivalent,
the quark CKM matrix). The theoretical Job 1 is to decipher this hidden
information (and similarly for quarks). After decades of world-wide
very hard work, no one suceeded in doing that as yet, at least according
to the opinion of the great majority of particle theoreticians.

The local probability for a neutrino, born at t=0 in flavour state$\beta$,
to survive in that flavour state at time t>0, at the spacetime point
labelled $\vec{x}=\vec{r}(t)$ is then according to elementary Quantum
Theory as formulated by the Bohm school

\begin{equation}
P_{\beta}(\vec{r}(t),t)=|\Psi_{\beta}(\vec{r}(t),t)|^{2}=\sum_{\alpha}P_{\beta\rightarrow\alpha}(\vec{r}(t),t)\qquad\qquad t\geq0\label{eq:90}\end{equation}

\begin{equation}
P_{\beta\rightarrow\alpha}(\vec{r}(t),t)=|\Phi_{\beta\alpha}(\vec{r}(t),t)|^{2}\qquad\qquad t\geq0\label{eq:91}\end{equation}

Recall that the history of our neutrino beable $\vec{r}_{n}(t)$ with
mass $mass\; m_{n}$ must be a solution of the Bohm equation (92)
with the 4-spinor guiding wavefunction given by (81) that satisfies
the appropriate initial condition:

\begin{equation}
\vec{v}(\{\vec{r}(t)\},t)=\{\frac{\Psi_{\beta}^{\dagger}(\vec{x},t)\vec{\alpha}\Psi_{\beta}(\vec{x},t)}{\Psi_{\beta}^{\dagger}(\vec{x},t)\Psi_{\beta}(\vec{x},t))}\}_{\vec{x}=\vec{r}(t)}\equiv\frac{d}{dt}\vec{r}(t)\qquad t\geq0\label{eq:92}\end{equation}

The flavour oscillations known to exist are therefore a direct result
of the interferences among sinusoidal terms originating in the guiding
$wavefunction$ (86) and that results from result (87) {[}22{]}. It
can be easily shown that they depend on the mass squared differences
of the three neutrino species. If the neutrinos were mass degenerate,
then there would be no oscillations, contrary to the experimental
results.

I would like to emphasize that from the Bohm-Hiley's viewpoint one
is really discussing here the probability that our neutrino beable
$is$ (and not merely {}``found if measured'' in textbook parlance!)
at the position $\vec{r}(t)$ $and$ with the velocity $\vec{v}(\vec{r}(t);t)$
when the time is t>0 , having started at time t=0 at a distance L
in that channel in some definite mean energy but given flavour $\beta$
with the probability $P_{\beta}(\vec{r}(0)=\vec{L},t=0)$ What is
thus interesting has very little to do with spin. On the other hand,
let us remind ourselves that $bona\; fide$ experiments are always
carried out with large ensembles of  particles; an {}``observation''
done with a single neutrino has no statistical significance at all! 

The intervention of any {}``flavour measuring device'' in the present
case has - according to textbooks at least - an inevitable consequence,
that is, it boils down - in some very unclear way - to an ill-defined
{}``collapse of the wavefunction'' into some definite flavour eigenstate
{[}3{]}. Then, in some vaguely specified way, this gives the Born
probability (90) that immediately after the {}``measurement is over''
the neutrino {}``will be found'' in some definite flavour state.

The Bohm school considers all this basically quite unsatisfactory.
As suggested above, the answers, suggestions and explanations given
by the Bohm school are thus totally different from those we are accustomed
to {[}2{]},{[}3{]}, and apply equally well to individual beable particles
and to ensembles.

So, there is no {}``collapse'' of any kind, as the wavefunction
is a beable just as a beable neutrino. Its guiding wavefunction, that
is a solution of the equation of motion (85), $actively$ $informs$
in principle the beable neutrino of all the infinite $potentialities$
of the Universe in which this particular neutrino can exist {[}2{]}{[}3{]}.
Any possible so-called {}``measurement'' that {}``reveals'' the
neutrino flavour state is basically in Bohm's view just an example
of a quantum transition, ultimately initiated by collisions with virtual
very heavy $W^{\pm}$electroweak vacuum boson modes {[}2{]}.

A more detailed and technical account of neutrino vacuuum flavour
oscillations along these lines will be given elsewhere {[}24{]}.

\subsection{Uniformly accelerated particle motions in vacuum }

According to the standard point of view of modern Particle Physics
we are living in a Minkowski universe. {}``We'' means here some
inertial frame, call it {}``the LAB (inertial) reference frame''.
Imagine that the LAB frame is watching an uniformly accelerated particle,
say a Ps atom. Consider then $co-moving$$\;$$inertial$ reference
frames, which are here named {}``BODY frames'', moving along with
the particle {[}25{]}. Thus at any definite proper time $\tau$ we
can apply in the BODY frame at that time the theory sketched in subsection
(3.1).

Let us now discuss this from the Bohm-Hiley version of RQFT, as interpreted
in the present paper (incidentally, a most lucid and pedagogical dicussion
of the physics of Lorentz boosts was given by the same authors {[}26{]}).

The questions will be translated in a more technical language: how
would the world look like from the point of view of any such BODY
frame? If the atom is not under accelerations, then the answer is
given in subsection 3.1. If the atom is in an uniformly accelerated
motion, then the answer can be easily found as shown by J.Donnoghue
and B.Holstein {[}27{]}, by simply transforming the metric tensor
from a LAB Minkowski frame to any co-moving BODY frame using a coordinate
system appropriate to a Rindler universe {[}25{]}.

The object that has our interest in this connection is the time-integrated
Minkowski correlation function closely related to the definitions
(71) and (72):

\[
\int_{0}^{t}dt_{2}\int_{0}^{t_{2}}dt_{1}G_{Mij}(\vec{x}_{1},t_{1};\vec{x}_{2},t_{2})\equiv\]

\[
\frac{1}{2!}\int_{0}^{t}dt_{2}\int_{0}^{t}dt_{1}\{<0_{M}|\Theta(t_{1}-t_{2})(A_{Ti}(\vec{x}_{1},t_{1})A_{Tj}(\vec{x}_{2},t_{2}))|0_{M}>+\]

\begin{equation}
+<0_{M}|\Theta(t_{2}-t_{1})(A_{Tj}(\vec{x}_{2},t_{2})A_{Ti}(\vec{x}_{1},t_{1}))|0_{M}>\}\label{eq:93}\end{equation}

where $\Theta$ is the usual step function (assuming that $\eta\longrightarrow0+)$: 

\begin{equation}
\Theta(\tau)\equiv\frac{-1}{2\pi i}\int_{-\infty}^{+\infty}\frac{d\omega}{\omega+i\eta}e^{-i\omega\tau}\label{eq:94}\end{equation}

|0$_{M}>$is supposed to mean {}`` the Minkowski vacuum in any inertial
reference frame''. So, by-passing irrelevant complications due to
sum over polarization vectors, one finds that {[}27{]}

\[
\int_{0}^{t}dt_{2}\int_{0}^{t_{2}}dt_{1}G_{Mij}(\vec{x}_{1}=0,t_{1};\vec{x}_{2}=0,t_{2})=\]

\begin{equation}
=2(\hbar c^{2})\int\frac{d^{3}\vec{k}}{(2\pi)^{3}2|\vec{k}|}exp[-i|\vec{k}]c(t_{2}-t_{1})]\label{eq:95}\end{equation}

We leave this integral in this form becuse we are concerned only with
its spacetime-dependence.

Then as proved in ref {[}27{]} much of the discussion on subsection
(3.1) can be directly adapted to a Rindler metric in any co-moving
BODY frame, as follows. Let us make a change from the Minkowski metric
to the Rindler metric, corresponding to a boost along the z-axis producing
an uniform acceleration $a$ with which the atom's center of mass
moves from the point of view of the LAB system:

\begin{equation}
x=y=0\qquad z=\frac{c^{2}}{a}(cosh\frac{a\tau}{c}-1)\qquad t=\frac{c}{a}sinh\frac{a\tau}{c}\label{eq:96}\end{equation}

\begin{equation}
c^{2}d\tau^{2}=c^{2}dt^{2}-dx^{2}-dy^{2}-dz^{2}\label{eq:97}\end{equation}

$\tau$ is the proper time registered by a co-moving observer. As
it is shown in {[}27{]} by simply making this transformation one finds
that the result (95) now becomes for a Rindler universe:

\[
\int_{0}^{t}dt_{2}\int_{0}^{t_{2}}dt_{1}G_{Rij}(\vec{x}_{1}=0,t_{1};\vec{x}_{2}=0,t_{2})=\]

\[
=2(\hbar c^{2})\int\frac{d^{3}\vec{k}}{(2\pi)^{3}2|\vec{k}|}exp\{-i|\vec{k}]c\tau\}\times[1+\frac{1}{exp(2\pi|\vec{k}|c/a)-1}]+\]

\begin{equation}
+2(\hbar c^{2})\int\frac{d^{3}\vec{k}}{(2\pi)^{3}2|\vec{k}|}\frac{exp\{i|\vec{k}]c\tau\}}{exp(2\pi|\vec{k}|c/a)-1}\label{eq:98}\end{equation}

This remarkable result reduces of course to the Minkowski result (95)
if the acceleration is set to zero.

As shown in {[}27{]} this is exactly the same result that co-moving
inertial observers would find, if the atom actually moved in a thermal
bath of temperature

\begin{equation}
T=\frac{a}{2\pi}\label{eq:99}\end{equation}

Summing up: from the point of view of a Minkowski LAB inertial system
the vacuum is a quantum medium that to a reasonable approximation
looks like an infinite sea of non-interacting quantum harmonic oscillators.
However, for a co-moving BODY reference system (i.e. within a Rindler
universe) the vacuum would look like a heath bath with the Unruh temperature
(99){[}27{]}.

\section{Summary and conclusions }

I have attempted to sketch and apply in this paper a personal interpretation
of the original Bohm project, in a manner that might be understandable
to dedicated members of the large community of Particle physicists
who may not already be familiar with it. This is simply because this
particular formulation seems to me to be the most powerful, natural
and credible one among its many concurrents.

After a short summary of some of the relevant background material
(Section 2) the attempt is made to $explain$ - i.e. in the Bohmean
sense of this word (Section 1) - three randomly chosen, but particularly
instructive, case studies (Section 3) borrowed from published papers
and textbooks on Particle Physics.

Bohm's opinion (see the quotation in Section 1) was apparently that
Quantum Theory in its present official (most would perhaps add - final?)
formulation, dating back to the thirties, $explains$ nothing. Nevertheless,
it does give an excellent, correct and precise qualitative and quantitative
$description$ of what any experimenter will see, or not see, in any
of his measuring apparatus, once the correct specifications are satisfied.

This is usually taken to be all there is to it.

I have argued throughout this paper for an alternative point of view
due to Bohm. One ought at present to play with both conceptions: the
{}``romantic'' textbook epistemological/pragmatical version and
by now the many alternative {}``reallistic'', or {}``business-like''
ontological versions {[}3{]}.

The particular Bohm ontological version is (according to this author's
taste at least) the most satisfying one among by now countless alternative
ontological versions. 

Those who are not entirely happy with the standard formulations of
Quantum Theory, nor e.g. with the Bohm version, will have to compromise,
until the Bohm causal formulation, or perhaps some alternative one,
become mature enough to stand on its own feet and ready to face head
on the inevitable experimental challenges of the coming decades.

\end{document}